\documentclass[apj,a4paper,12pt,useAMS]{emulateapj}

\usepackage{amsmath}
\usepackage{cases}
\usepackage{amssymb}
\usepackage{graphicx}

\begin{document}
%\linenumbers

%\title{could proto-Quark Star be the central engine of Gamma Ray Bursts ?}
%\title{The early stage thermal evolution and The Surface $e^{\pm}$  Wind Emission from Bare proto-Quark Stars}
\title{A powerful $e^{\pm}$ outflow driven by a proto-strange quark star}
\author{Shao-Ze Li$^{1}$, Yun-Wei Yu$^{2}$, He Gao$^{1}$, Zi-Gao Dai$^{3,4}$, and Xiao-Ping Zheng$^{2}$}

\altaffiltext{1}{Department of Astronomy, Beijing Normal University,
Beijing 100875, China; gaohe@bnu.edu.cn} \altaffiltext{2}{Institute
of Astrophysics, Central China Normal University, Wuhan 430079,
China; yuyw@mail.ccnu.edu.cn} \altaffiltext{3} {School of Astronomy
and Space Sciences, University of Science and Technology of China,
Hefei 230026, China} \altaffiltext{4} {School of Astronomy and Space
Science, Nanjing University, Nanjing 210023, China}

\begin{abstract}
An electron-positron layer can cover the surface of a bare strange
star (SS), the electric field in which can excite the vacuum and
drive a pair wind by taking away the heat of the star. In order to
investigate the pair emission ability of a proto-SS, we establish a
toy model to describe its early thermal evolution, where the initial
trapping of neutrinos is specially taken into account. It is found
that the early cooling of the SS is dominated by the neutrino
diffusion rather than the conventional Urca processes, which leads
to the appearance of an initial temperature plateau. During this
plateau phase, the surface $e^{\pm}$ pair emission can keep a
constant luminosity of $10^{48}-10^{50}\rm erg\,s^{-1}$ for about a
few to a few tens of seconds, which is dependent on the value of the
initial temperature. The total energy released through this
$e^{\pm}$ wind can reach as high as $\sim10^{51}\rm erg$. In
principle, this pair wind could be responsible for the prompt
emission or extended emission of some gamma-ray bursts.
\end{abstract}

\section{Introduction}
It was hypothesized that strange quark matter (SQM), consisting of
roughly equal numbers of up, down, and strange quarks, is the ground
state of strong interaction (Witten 1984). If this hypothesis is
true, then the nature of the compact stars known as pulsars should
in fact be strange quark stars (SSs; Alcock et al. 1986; Haensel et
al. 1986) rather than conventional neutron stars (NSs), at least in
part. Nevertheless, the existence of NSs can still be allowed as a
metastable state. It is undoubtedly of fundamental physical
significance to test the SS hypothesis by searching for possible
candidates from observations. According to the differences between
SSs and NSs in their mass-radius relations, cooling histories, and
rotational properties, several observational signatures have been
proposed for identifying SS candidates (e.g. Pizzochero 1991; Page
1992; Schaab et al. 1996; Madsen 2000; Glendenning et al. 2001; Xu
et al. 2002; Ray et al. 2004; Bagchi et al. 2006; Yu \& Zheng 2006;
Zheng et al. 2006). Unfortunately, current observations of Galactic
pulsars usually found them falling into an ambiguous parameter space
that covers both predictions of the SS and NS models (Lorimer \&
Kramer 2012; Lyne \& Graham-Smith 2012).

During the past two decades, it has been increasingly suggested that
a rapidly rotating and highly magnetized pulsar can be formed from
some extragalactic luminous transient phenomena such as gamma-ray
bursts (GRBs; Dai \& Lu 1998a,b; Dai et al. 2006; Yu et al. 2010)
and superluminous supernovae (SLSNe; Woosley 2010; Kasen \& Bildsten
2010; Dexter \& Kasen 2013). The substantial influence of the
newborn pulsars on the transient emission could provide
a potentially effective way to probe into the properties and nature of these
compact stars (see Yu et al. 2019 for a brief review). For example,
SSs are expected to be able to keep near-Keplerian rotation much longer
than NSs due to the high bulk viscosity of SQM. Then, Yu et al. (2009)
suggested that SS candidates could be picked up from the
afterglow observations of GRBs by searching for ultra-long
internal plateau afterglows, since these plateau emission is very likely to be powered by the stellar rotation.
Following this consideration, Dai et al. (2016) further suggested that the
central engine of the most luminous supernova ASASSN-15lh could just be
an extremely rotating SS.

More excitingly, mergers of double pulsars, which can result in a
short GRB, could give birth to a new massive pulsar, which is very
helpful for understanding the afterglow features of short GRBs (Dai
et al. 2006; Rowlinson et al. 2010, 2013; Bucciantini et al. 2012;
Gompertz et al. 2013, 2015; Zhang 2013; L\"{u} et al. 2015; Gao et
al. 2015). Some new clues to such a post-merger pulsar could have
also appeared in the multi-messenger observations of GW events
(e.g., GW170817), such as kilonova (Yu et al. 2018; Li et al. 2018;
Ai et al. 2018; Radice et al. 2018; Piro et al. 2019; Margutti et
al. 2019; Troja et al. 2020; Ren et al. 2020). Then, in comparison
to the pulsars originating from core collapse of massive stars, we
more expect that the nature of post-merger pulsars are inclined to
be SSs. Even if the pre-merger pulsars are NSs, a phase transition
from metastable NSs before merger to an SS after merger could still
happen, in view of the high mass $\gtrsim2.0M_{\odot}$ of the merger
product (e.g., Alford et al. 2019; Parisi et al. 2020). Such an
expectation can indeed be supported by the fittings of the
afterglows of some short GRBs (Li et al. 2016; Hou et al. 2018). In
principle, the most direct signature of a post-merger SS can be
digged from the waveform of the GW signal (Oechslin et al. 2004;
Bauswein et al. 2010), which is however far beyond the ability of
current GW detectors (Bauswein et al. 2019; Most et al. 2020).
Therefore, at present, it is still necessary to investigate the
other astrophysical manifestations of newborn SSs, which acctaully
has been long concerned by the community (Dai et al. 1995; 2016;
Cheng \& Dai 1998;
 Ouyed et al. 2002; 2009; 2015; Leahy \& Ouyed 2008; Cheng et al. 2009a; 2009b;
  Cheng \& Harko 2010; Pagliara et al.2013).

Due to the extremely high temperature at birth, a newborn pulsar
cannot have a solid crust, which enables SSs show a special feature.
To be specific, different from NSs, the surface of bare SSs should
be covered by an electron layer of a thickness of a few thousand of
femtometres\footnote{As the stellar temperature decreases to the
melting point, a thin crust consisting of electrons and ions but
without free neutrons can be supported by the electron layer (Alcock
et al. 1986; Zheng \& Yu 2006), which makes the surface of SSs be
similar to that of NSs.} (Alcock et al. 1986), because of the
different ranges of the strong force between quarks and the
electromagnetic force between quarks and electrons. This separation
of positive and negative charges generates a radially outwards
electric field of a typical magnitude of $\sim10^{17}\rm V~cm^{-1}$
(Alcock et al. 1986; Zheng \& Yu 2006). Such an extremely high
electric field, wihch is a few to a few tens of times higher than
the critical field $E_{\rm c}=m_{\rm
e}^2c^3/e\hbar=1.3\times10^{16}\rm  V~cm^{-1}$,  can excite the
vacuum to create $e^{\pm}$ pairs (Schwinger 1951). Therefore, Usov
(1998; 2001) suggested that a remarkable number of $e^{\pm}$ pairs
can be produced from the electron layer of a bare SS (i.e., Usov
mechanism), which leads to an $e^{\pm}$ wind by taking away the heat
of the SS. The luminosity of this wind can be estimated to $10^{40}-
10^{44}\rm erg\, s^{-1}$ for typical temperatures of young pulsars
(Page and Usov 2002). Furthermore, if the wind can effectively
absorb energy from neutrinos, it could further be accelerated into a
relativistic speed (Cheng et al. 2009b; Cheng \& Harko 2010).

The purpose of this paper is to describe the surface
$e^{\pm}$ emission of proto-SSs, by according to the modeling of their initial cooling
processes. The results could help us to understand the possible
astrophysical consequences of the formation of proto-SSs and, then, to identify
them.
%In section 2, we build a simple model to describe the neutrino
%diffusion process in case of that most of the neutrinos are trapped
%inside the proto-SS initially. Using this diffusion model, we could
%obtain the effective temperature at the neutrino sphere, which
%allows us to calculate the surface $e^{\pm}$ emission. The results
%are shown in section 3 and the conclusion and discussion is in
%section 4.

%\textbf{In section 2, we discuss the neutrino trapping and the
%energy deposition in proto SSs. In section 3, we build a simple
%diffusion model to describe the early thermal evolution. Using this
%diffusion model, we could obtain the effective temperature at the
%neutrino sphere, which allow us to calculate the surface $e^{\pm}$
%emission. The results are shown in section 4 and the conclusion and
%discussion is in section 5.}
\section{The model}
\subsection{Basic equations}
When an SS is born, its initial temperature can be estimated to be
as high as several tens of MeV, by supposing half of the
gravitational energy of the stellar material has been converted into
the internal energy. Furthermore, the phase transition from baryons to quarks as a robust
deflagration can also release a great amount of energy in a short
timescale (Pagliara et al. 2013), which can in principle lead
proto-SSs to be much hotter than proto-NSs.
%Similarly with proto-NSs, the proto-SSs quickly lose their energy through neutrino emission, which can reach as high as $10^{53}\,\rm erg\,s^{-1}$ (Pons et al. 2001; Pagliara et al. 2013).
The internal energy of a proto-SS is shared by
quarks, electrons, photons, and neutrinos. Then, the SS can
be cooled via the escaping of neutrinos from the stellar
interior and the escaping of photons and $e^{\pm}$ pairs from the
stellar surface. Then, by denoting the total internal energy of the SS as $\mathcal E$, its evolution can be determined by
\begin{eqnarray}
{d\mathcal E\over d t}=-L_{\nu}-L_{\gamma} -L_{\pm},\label{cooleq1}
\end{eqnarray}
where $L_{\nu}$, $L_{\gamma}$, and $L_{\pm}$ are the luminosities of
the neutrino, photon, and pair emission,
respectively. Specifically, the luminosity of the pair emission due to the Usov mechanism is given by
\begin{eqnarray}
L_{\pm}\approx4\pi
R^{2}\varepsilon_{\pm}\dot{n}_{\pm},\label{Lpair}
\end{eqnarray}
where $R$ is the stellar radius, $\varepsilon_{\pm}=m_{e}c^{2}+kT_{\rm s}$ is the average energy of leptons with $T_{\rm s}$
being the surface temperature, and $\dot{n}_{\pm}$ is the number flux which reads (Usov
1998; 2001)
\begin{eqnarray}
\dot{n}_{\pm}\approx10^{42}T_{\rm s,10}^{3}{\rm exp}\left(-{1.2\over T_{\rm s,10}}\right)\,J(\zeta)\,s^{-1},\label{npair}
\end{eqnarray}
where $\zeta=2({\alpha/\pi})^{1/2}{\varepsilon_{\rm F,l}/( k_{\rm
B}T_{\rm s}})$, $\alpha$ is the fine-structure constant,
$\varepsilon_{\rm F,l}$ is the Fermi energy of leptons, and
\begin{eqnarray}
J(\zeta)={1\over3}{\zeta^{3}{\rm
ln}(1+2\zeta^{-1})\over(1+0.074\zeta)^3} +
{\pi^5\over6}{\zeta^4\over(13.9+\zeta)^4}.
\end{eqnarray}
Hereafter the conventional notation $Q_x=Q/10^{x}$ is used in the
cgs units. Meanwhile, one may write the luminosity of the surface photon emission by using the Stephan-Boltzmann formula. However, as pointed out by Usov (2001), the black body emission can actually be significantly suppressed for $T_{\rm s}>8\times
10^{8}\rm K$, since the diffusion of the photons of an electromagnetic frequency lower than the plasma
frequency of the quark matter can be blocked. Therefore, in our calculations we can simply
neglect $L_{\gamma}$.

For the neutrinos, if they can escape freely as considered for a
normal SS, then the luminosity can usually be calculated by timing
the stellar volume to the emissivity $\epsilon_\nu$ (Iwamoto 1980)
\begin{eqnarray}
L_{\nu,\rm free}&=&{4\over3}\pi R^3 \epsilon_\nu\nonumber\\
 &\approx& 3.7\times10^{57}R_{\rm 6}^3\alpha_{\rm c}\left({n_{\rm b}\over
n_{\rm 0}}\right)Y_{e}^{1/3}T_{11}^{6}\rm erg~s^{-1}, \label{nulum}
\end{eqnarray}
which is obtained from the direct Urca process $d\rightarrow u e^{-}
\bar{\nu}_{e}$ and $u  e^{-} \rightarrow d   \nu_{e}$, where
$\alpha_{\rm c}=0.1$ is the strong coupling coefficient, $n_{\rm b}$
is the baryon number density, $n_{\rm 0}=0.16\rm fm^{-3}$ is the
nuclear saturation density, $Y_{\rm e}$ is the electron fraction in
the interior of the SS, and $T$ is the inner temperature.
Nevertheless, it should be pointed out that Equation (\ref{nulum})
is actually inapplicable for a proto-SS which has an extremely
high temperature. In this case, the mean free path of neutrinos
due to the absorbtion and scattering by the quark matter can be much
shorter than the stellar radius, which reads (Iwamoto 1982;
Berdermann et al. 2004)
\begin{eqnarray}
l_{\nu}&\approx& 1.4\times 10^{3}\left({n_{\rm b}\over
n_{0}}\right)^{-1} Y_{e,-1}^{-1/3}\left({\mu_{e}\over
\mu_{\nu}}\right)^{-2}\nonumber\\
& & \left[1+{1\over 2}\left(\mu_{e}\over \mu_{u}\right) + {1\over
10}\left( {\mu_{e} \over \mu_{u}}
\right)^{2} \right]^{-1}\nonumber\\
& &\left[{\left(\varepsilon-\mu_{\nu} \over 10{\rm MeV}\right)^2}
+ {\left(\pi k T \over 10{\rm MeV}\right)^2}\right]^{-1}{\rm cm}
%\nonumber\\
%&\approx& 93\nonumber\\
%& &\left({n_{\rm b}\over n_{0}}\right)^{-1} \left({\mu_{e}\over
%\mu_{\nu}}\right)^{-2} \left({Y_{e}\over
%0.1}\right)^{-1/3}\left({kT\over {10\rm MeV}}\right)^{-2}{\rm
%cm}
\label{freepath}
\end{eqnarray}
For a rough estimation, the neutrino energy can be adopted as
$(\varepsilon-\mu_{\nu})\sim kT$ and the chemical potentials satisfy
$\mu_{e}\sim \mu_{\nu} <  \mu_{u}$. This indicates the neutrinos can
be seriously trapped in the proto-SS until the stellar temperature
decreases to be lower than $\sim 10^{9}$ K.

\begin{figure}[t]
\centering\resizebox{1.0\hsize}{!}{\includegraphics{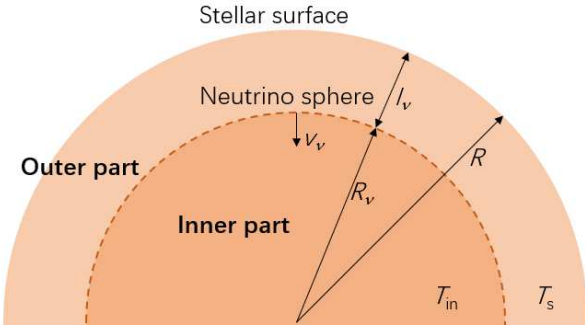}}\caption{An illustration of the model, where the proto-SS is divided into an inner and outer part by the neutrino sphere.}\label{illustration}
\end{figure}
Therefore, for the
initial cooling of the proto-SS, the neutrino luminosity should be
alternatively recalculated by considering of the neutrino diffusion
in the stellar interior. For an approximative calculation of the neutrino emission, it could be convenient to
separate the proto-SS into two parts by according to a neutrino sphere at the radius of
\begin{eqnarray}
R_{\nu}=R-l_{\nu},
\end{eqnarray}
as illustrated in Figure \ref{illustration}. %Outside which the neutrinos can escape freely. %On the contrary, for the neutrinos inside the neutrino sphere, they must undergo a diffusion process first.
Furthermore, the temperatures in the inner and outer parts are both considered to be uniform, which are referred to the effective internal and
surface temperature, respectively. The actual temperature gradient
in the stellar interior is simplified by a temperature jump
occurring at the neutrino sphere. Then, the neutrino luminosity at
$R_{\nu}$ is determined by the neutrino diffusion from the stellar
center to the neutrino sphere and, for an order-of-magnitude
estimation, the luminosity can be given by (Kasen \& Bildsten 2010;
Yu et al. 2015)
\begin{eqnarray}
L_{\nu,\rm in}={4\pi R_{\nu}^{2}c\over 3\kappa_{\nu}\rho}{\partial
u_{\nu}\over\partial r}\sim{4\pi R_{\nu}^{2}c(u_{\nu,\rm
in}-u_{\nu,\rm out})\over 3\max(\tau_{\nu},1)},
\end{eqnarray}
where $u_{\nu}$ is the energy density of thermal neutrinos, $\kappa_{\nu}$ is
the opacity, $\rho$ is the mass density, and
$\tau_{\nu}=\kappa_{\nu}\rho R_{\nu}=R_\nu/l_{\nu}$ is the neutrino
optical depth of the inner part. Meanwhile, the neutrino
emission of the outer part of the star is contributed by two
components, including the escaping of the thermal neutrinos and the
neutrino production via the Urca processes. Then, the luminosity of
outer part can be written as
\begin{eqnarray}
L_{\nu,\rm out}&=& \pi R^2cu_{\nu,\rm out}+{4\over3}\pi
(R^3-R_{\nu}^3)\epsilon_\nu.
\end{eqnarray}
Here the appearance of the thermal emission component is due to the
absorption of the neutrinos from the inner part by the outer part.

Following the above consideration, we further denote the internal
energy of the inner and outer part of the proto-SS by
\begin{eqnarray}
\mathcal E_{\rm in}={4\over3}\pi R_{\nu}^3u_{\rm in}
\end{eqnarray}
and
\begin{eqnarray}
\mathcal E_{\rm out}={4\over3}\pi\left(R^3- R_{\nu}^3\right)u_{\rm out}
\end{eqnarray}
respectively, where the energy density are given by summarizing all
particles as $u_{\rm }=\sum\limits_{i=q,e,\nu}u_i$.
%and $u_{\rm out}=\sum\limits_{i=q,e}u_i$?????.
Then, Equation (\ref{cooleq1})
can be replaced by the following two equations:
\begin{eqnarray}
{d\mathcal E_{\rm in}\over d t}&=&- L_{\nu,\rm in} -4\pi R_{\nu}^{2}
v_{\nu}u_{\rm in} \label{cooleqA}
\end{eqnarray}
and
\begin{eqnarray}
{d\mathcal E_{\rm out}\over d t}&=&\zeta L_{\nu,\rm in}+4\pi
R_{\nu}^{2}v_{\nu }u_{\rm in} -L_{\nu,\rm out}-L_{\pm},
\label{cooleqB}
\end{eqnarray}
where $\zeta=1-e^{-1}=0.63$ represents the absorption by the neutrino sphere, the term $4\pi R_{\nu}^{2}v_{\nu}u_{\rm in}$ represents the
energy transfer from the inner to the outer part due to the inward
movement of the neutrino sphere, and $v_{\nu}={-dR_{\nu
}/dt}=dl_{\nu}/dt$ is the moving velocity of the neutrino sphere. Finally, the total luminosity of the neutrino emission is given by
\begin{eqnarray}
L_\nu=(1-\zeta)L_{\nu,\rm in}+L_{\nu,\rm out}.
\label{}
\end{eqnarray}
In our model, we do not consider the possible change of the
stellar radius due to the change of the equation of state during the
phase transition, which could only lead to  a $\sim1\%$ difference
(Dexheimer et al. 2013).

\subsection{Expressions of internal energy}
The temperature
dependence of the internal energy of the SQM is determined by
the $\beta$ equilibrium in presence of neutrinos: $u + e^{-} \rightleftharpoons  d + \nu$ and
$u + e^{-} \rightleftharpoons  s + \nu$. In this neutrino trapping
case, the chemical potentials of the particles are suggested to
approximately satisfy (Iwamoto 1982)
\begin{eqnarray}
\mu_{ u} \simeq \mu_{ d} \simeq \mu_{ s}
\end{eqnarray}
and
\begin{eqnarray}
\mu_{ e} \simeq \mu_{\nu}.
\end{eqnarray}
Then, for
relativistic quarks, their internal energy can be calculated as (Iwamoto 1982; Shapiro \& Teukolsky 1983),
\begin{eqnarray}
u_{\rm q}&=&\sum \limits_{i=u,d,s}{\pi^2 \over
2}n_{i}\varepsilon_{{\rm F},i}\left({k_{\rm
} T\over \varepsilon_{{\rm F},i}}\right)^2\nonumber \\
&\approx&1.7\times 10^{33}\left({n_{\rm b}\over
n_{0}}\right)^{2/3}\left({k_{\rm }T\over 10 \rm MeV}\right)^{2}\rm
erg\,cm^{-3}\label{uq}  %%2-24这个值要check一下
\end{eqnarray}
where the Fermi energy of the quarks reads
\begin{eqnarray}
\varepsilon_{{\rm F},\rm q} \simeq \left({n_{\rm b}h^3c^3
\over8\pi}\right)^{1/3} \approx 235\left({n_{\rm b}\over
n_{0}}\right)^{1/3}\rm MeV
\end{eqnarray}
for $n_{\rm u}=n_{\rm d}=n_{\rm s}=n_{\rm b}$. Here, because the
baryon number is conserved, the degeneracy energy of the quarks
cannot be released during the stellar cooling. Thus, only the
thermal component is necessary and exhibited in Eq. (\ref{uq}).
Different from quarks, neutrinos can eventually escape from the proto-SS.
As a result, the number density and chemical potential of electrons
and neutrinos can decrease with time. Therefore, when we describe the internal energy of the leptons, both the thermal and degeneracy components should be taken into account as
\begin{eqnarray}
u_{\rm l}&=&\sum
\limits_{i=e,\nu}{\left[{3\over4}n_{i}\varepsilon_{{\rm F},i}+{\pi^2
\over 2} n_{i}\varepsilon_{{\rm F},i}\left({k_{\rm
} T\over \varepsilon_{{\rm F},i}}\right)^2 \right]}\nonumber\\
&\simeq&{{4\pi
 \over c^3 h^3}\mu_{\rm l}^4\left[1+{2\pi^2
 \over 3 }\left({k_{\rm
} T\over \mu_{\rm l}}\right)^2\right]}, \label{uv}
\end{eqnarray}
where $\mu_{\rm l}=\varepsilon_{\rm F,l}$ is used. The above expression is valid for $kT\ll \varepsilon_{\rm F,l}$,
where the Fermi energy of the leptons can be expressed by introducing an electron fraction $Y_{\rm e}$ as
\begin{eqnarray}
\varepsilon_{\rm F,l}&=&(3Y_{\rm e})^{1/3}\varepsilon_{\rm F,q}\nonumber\\
&\approx&200 \left(Y_{\rm e}\over 0.2\right)^{1/3}\left(n_{\rm
b}\over n_{0}\right)^{1/3}\rm
 MeV.
\end{eqnarray}

As the escaping of the neutrinos, the chemical equilibrium in the
outer part finally becomes
\begin{eqnarray}
\mu_{ u}+\mu_{ e} \simeq \mu_{ d} \simeq \mu_{ s}.
\end{eqnarray}
Here the neutrino chemical potential $\mu_\nu$ is believed to be
zero, although it is not easy to explicitly describe the decreasing
behavior of $\mu_\nu$, which is non-equilibrium. Then, for an
effective description of the energy density of neutrinos in the
outer part, it is assumed that the neutrino chemical potential can
simply co-evolve with the temperature as\footnote{In the case of
that the free path $l_{\nu}$ is not sensitive to $Y_{e}$ as in Eq.
6, so $l_{\nu}/R\simeq T^{2}_{\rm cr}/T^{2}$ is used in our
calculation for simplicity, where $T_{\rm cr}$ is the critical
temperature when the SS become transparent to neutrinos.}
\begin{eqnarray}
\mu_{\nu}\propto k_{\rm }T e^{-l_{\nu}/R},
\end{eqnarray}
which is basically in agreement with the
considerations in previous works (e.g. Pons et al. 1999; 2001;
Benvenuto \& Horvath 2013; Pagliara et al.2013). Here the term $
e^{-l_{\nu}/R}$ is introduced by considering that the neutrino
chemical potential should quickly approach to be zero after the SS
becomes totally transparent for $l_{\nu}>R$. In addition, a constant
value of $10^{-3}$ will be assigned to the electron faction in the outer
part, after the electrons are finally decoupled with the
neutrinos.

\begin{figure}[t]
\centering\resizebox{1.0\hsize}{!}{\includegraphics{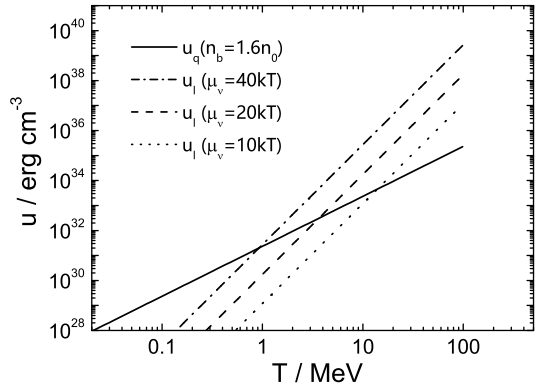}}\caption{The
energy density of quarks and neutrinos as functions of temperature.
The chemical potential of the neutrinos is assumed to be
proportional to the temperature and three tentative values are taken
for the proportional coefficient. }\label{energy density}
\end{figure}

In Fig. \ref{energy density}, we present a comparison between the
temperature dependence of $u_{\rm q}$ and $u_{\nu}$. For a
relatively high temperature, the internal energy is primarily stored
in the neutrinos and thus the cooling of the proto-SS should be
dominated by the release of the thermal neutrinos. On the contrary,
for a relatively low temperature, the neutrino trapping is relaxed
and the quark energy becomes dominated. In this case, the Urca
processes are the primary channel of the stellar cooling.

\section{results and discussion}\label{results}

\begin{figure}[t]
\centering\resizebox{1.0\hsize}{!}{\includegraphics{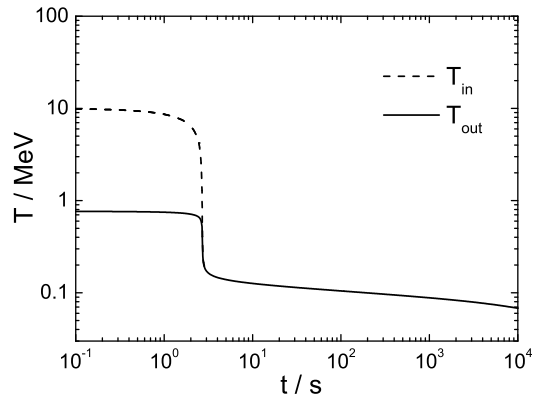}}\caption{The
temporal evolution of the internal and surface temperatures of an
SS. The initial temperature $T_{\rm i}=10 \rm MeV$ and electron
fraction $Y_{\rm e}=0.2$. The baryonic mass and radius of the SS are
taken as $1.56 M_{\odot}$ and $12\rm km$, respectively, which
determines an uniform density of $n_{\rm b}=1.6 n_{0}$.
}\label{cooling}
\end{figure}

The internal energy of a proto-SS is initially converted
from the gravitational energy released during the formation of the
SS and from the phase transition from nucleon to quarks.
Then the initial temperature can be estimated to be on the order of magnitude of $kT_{\rm i}\sim 10$ MeV by according to
\begin{eqnarray}
{4\over3}\pi R^3\left[u_{\rm q}(T_{\rm i})+u_{\rm l}(T_{\rm
i})\right]\sim\left( {3GM^2\over 10R}+{M\over m_{\rm p}}q \right),
\end{eqnarray}
where $q\approx 20$ MeV is the latent heat per baryon. For such an
extremely high temperature, the neutrino free path is much smaller
than the stellar radius and thus we take $R_{\nu}\approx R$ in the
above estimation. Then, for a reference value of $kT_{\rm i}=10$
MeV, we solve Equations (\ref{cooleqA}) and (\ref{cooleqB}) and
obtain the cooling curves of the SS as presented in Figure
\ref{cooling}, where both the internal and surface temperatures are
shown. Obviously, the cooling curves can be separated into a plateau
phase and a decline phase. Just as expected, the plateau is
determined by the neutrino diffusion, while the slow decay is
gradually dominated by the Urca processes.

\begin{figure}[t]
\centering\resizebox{1.0\hsize}{!}{\includegraphics{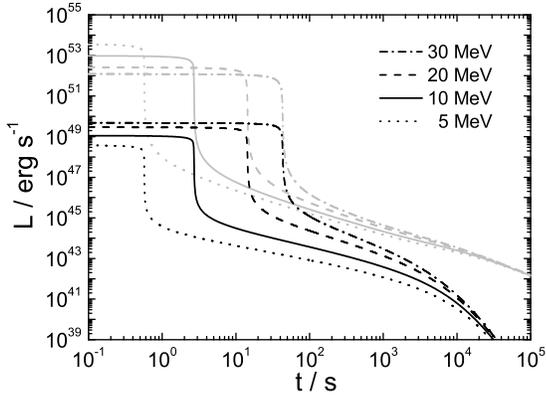}}\caption{The
luminosities of the $e^{\pm}$ pair emission (black) in comparison
with the neutrino luminosity (grey) from proto-SSs of different
initial temperatures. }\label{Lpairt}
\end{figure}

In Figure \ref{Lpairt}, we further present the evolution of the luminosity of the neutrino and pair emissions, for four different initial temperatures.
On the one hand, the
obtained neutrino luminosity are basically
consistent with the previous works (e.g. Pons et al. 2001; Pagliara
et al. 2013), which can be estimated by
\begin{eqnarray}
L_{\nu,\rm i}\approx {4\over3}\pi R^3{u_{\nu}\over t_{\rm d,\nu}} \sim10^{53}\rm erg\,s^{-1},
\end{eqnarray}
with a neutrino diffusion timescale of
\begin{eqnarray}
t_{\rm d,
\nu}=(R/l_{\nu})^2(l_{\nu}/c)\propto l_{\nu}^{-1}.\label{td}
\end{eqnarray}
This diffusion timescale determines the duration of
the plateau. According to Equation (\ref{freepath}), it can be
approximately obtained that $t_{\rm d, \nu}\propto T^2$. So, as
shown, the higher the initial temperature, the longer the plateau
duration. On the other hand, after the neutrino trapping, as a
result of the Urca process, the cooling for different initial
temperatures would gradually approach to be a common behavior. For
$L_{\nu}\propto T^6$, we can easily get $L_{\nu}\propto t^{-3/2}$
and $T_{}\propto t^{-1/4}$. During the plateau phase, the surface
temperature is much lower than the internal one, because the
internal energy cannot be effectively transported to the surface via
neutrino diffusion. This temperature gap can disappear as the
neutrino trapping is gradually relaxed. Nevertheless, for a
realistic consideration, a solid envelope could in principle be
formed as the temperature decreases to $<10^9$ K, if a sufficient
amount of material can be accreted onto the SS. In this case, a
significant temperature gradient can appear again in the envelope
because of its low thermal conductivity, which is beyond the scope
of this paper.

Finally, as a result of the cooling, we can obtain the evolution of
the luminosity of the surface $e^{\pm}$ pair emission, which may
have the most observational relevance. The pair luminosity during
the plateau phase can be as high as $\sim10^{49}-10^{50}\rm
erg~s^{-1}$. Different from the neutrino luminosity, the pair
luminosity is positively correlated with the initial temperature.
So, for a sufficiently high initial temperature, the total energy
taken away by the pair emission can even reach to $\sim10^{51}\rm
erg$, although the majority energy of the SS is released through the
neutrino emission.

%%不要用plateau这个词
%%从时标和光度上来看，我们的结果与simulation结果大体一致

%%！！！ 下面这句话，希望留着。因为我接下来的工作很可能是与protoNS与normalNS的个头大小不同相关的

%%Firstly, a great amount of internal energy would make significant
%%contribution to pressure. Secondly, the existence of degenerate
%%leptons change the beta balance and contributes additional
%%degenerate pressure. As a result, the EoS of proto-SS is dynamically
%%changing during early evolution.
%%The radius of proto SS may be a little different with normal
%%one. When thermal energy has been taken away, the radius returns
%%back to normal size.

%Different to proto-NS, %which has a very thick nuclear matter envelope,
%A bare proto-QUARK STAR is entirely composed of quark mater. Color
%interaction would keep the stellar not changing so much. From the
%calculation of Dexheimer et al. 2013, it has only about $1\%$
%difference. So it is suitable to assume a constant radius during the
%evolution.

%The result is basically consistent with the other works (e.g., Page
%and Usov (2001).

\section{Summary and Conclusion}

In this paper, we establish a toy model to calculate the thermal
evolution of a proto-SS, where the neutrino diffusion in the star is
described effectively by calculating the inward movement of the
neutrino sphere. This neutrino trapping effect has not be treated
seriously in previous works on the same topic (Page and Usov 2002;
Cheng et al. 2009b; 2010), which leads to a relatively high neutrino
luminosity at the beginning. Because of the neutrino trapping, the
early cooling of the SS is dominated by the neutrino diffusion
rather than the Urca processes. This leads to a temperature plateau
at the initial time, the duration of which is determined by the
timescale of the neutrino diffusion. As a result, although the
neutrino emission can take away the majority of the internal energy
of the SS, the surface $e^{\pm}$ pair emission due to the Usov
mechanism can still keep a constant luminosity of
$10^{48}-10^{50}\rm erg\,s^{-1}$ for about a few to a few tens of
seconds, which is dependent on the value of the initial temperature.
For a relatively high temperature, the total energy carried by the
$e^{\pm}$ wind can reach to be as high as $\sim10^{51}\rm erg$,
which seems high enough to cause some observable consequences.

For example, it was suggested that such an $e^{\pm}$ wind could give
rise to a GRB emission, if this wind can further be accelerated to
be relativistic by absorbing energy from the subsequent neutrino
flux. As mentioned in the introduction, the afterglow emission of
GRBs robustly suggest that some post-burst remnants could be a
pulsar rather a black hole. However, some people concerned that how
such a pulsar can produce a GRB outflow. It is even argued that only
black holes can generate a GRB. Then, the result obtained here
demonstrates that the $e^{\pm}$ wind from a bare proto-SS could be,
at least partially, responsible for the GRB trigger. To say the
least, if the GRB prompt emission indeed requires a hyper-accretion
onto the SS as suggested by Zhang \& Dai (2008a,2008b), the
$e^{\pm}$ emission of a constant luminosity for a few tens of
seconds could still provide a natural explanation for an extended
emission after the prompt emission, which is usually found in short
GRBs (Zhang 2013; L\"{u} and Zhang 2014; L\"{u} et al. 2015; Sun et
al. 2017). This may indicate that these short GRBs originate from a
merger of double pulsars and the merger product is a massive SS.

%Harko and Cheng (2006) calculated the $e^{\pm}$ emission rate
%carefully and show that a strong magnetic field at the bare quark
%star surface can significantly enhance the $e^{\pm}$ emission.

\acknowledgements  This work is supported by the National SKA
Program of China (grant No. 2020SKA0120300), the National Natural
Science Foundation of China (grant Nos. 11822302, 11833003,
11722324, 11690024, and 11633001), the Strategic Priority Research
Program of the Chinese Academy of Sciences (grant No. XDB23040100)
and the National Key Research and Development Program of China
(grant No. 2017YFA0402600) .

\end{document}